\documentclass[]{revtex4}
\usepackage{amsmath}
\usepackage{amssymb}
\usepackage{graphicx}
\usepackage{bm}

\def\degC{\,^\circ{\rm C}}
\def\cE{{\cal E}}
\def\cD{{\cal D}}

\def\eqnn#1{Eq.~(\ref{eq:#1})}
\def\sect#1{Sect.~\ref{sec:#1}}

\def\figno#1{Fig.~\ref{fig:#1}}

\def\tM{{T_{\rm meas}}} 

\def\vev#1{\langle #1\rangle}

\def\Re{{\rm \,Re}}

\def\cP{{\cal P}}

\def\dip{\mu_{\text Rb}}
\begin{document}
\title{
  Observing random walks of atoms  in buffer gas through resonant light absorption
}
\author{Kenichiro Aoki\footnote{E--mail:~{\tt ken@phys-h.keio.ac.jp}.}
  and Takahisa   Mitsui\footnote{E--mail:~{\tt      mitsui@phys-h.keio.ac.jp}.}}
\affiliation{Research and Education Center for Natural Sciences and
  Dept. of Physics, Hiyoshi, Keio University, Yokohama 223--8521,
  Japan}
\begin{abstract}
    Using resonant light absorption, random walk motions of rubidium
    atoms in nitrogen buffer gas are observed directly.  The
    transmitted light intensity through atomic vapor is measured and
    its spectrum is obtained, down to orders of magnitude below the
    shot noise level to detect fluctuations caused by atomic
    motions. To understand the measured spectra, the spectrum for
    atoms performing random walks in a gaussian light beam is computed
    and its analytical form is obtained. The spectrum has $1/f^2$
    ($f$: frequency) behavior at higher frequencies, crossing over to
    a different, but well defined behavior at lower frequencies.  The
    properties of this theoretical spectrum agree excellently with the
    measured spectrum. This understanding also enables us to obtain the
    diffusion constant, the photon cross section of atoms in buffer
    gas and the atomic number density, from a single spectral
    measurement.  We further discuss other possible applications of our
    experimental method and analysis.
\end{abstract}
\maketitle
\section{Introduction}
\label{sec:intro}
Thermal motion is inevitable for any object at finite temperatures. On
the microscopic scale, perhaps the most well known example is the
Brownian motion of particles, which is optically
visible\cite{Brown0,Einstein,Brown1}.  Even on a smaller scale, the
thermal motions of the atoms are visible through surface fluctuations
of liquids\cite{ripplon,ripplonExp,giant}, high power interferometry
measurements on mirrors\cite{mirror} and complex materials\cite{am1}.
In gases, the ballistic thermal motion of atoms and molecules lead to
the transit time broadening of the resonant
widths\cite{transitTimeBroadening} and the free streaming of atoms
can be observed through their transit noise in light\cite{amRb}.
When atoms have relatively shorter mean free paths, such as when
buffer gas is present, we expect the atoms to perform random walk
behavior caused by collisions with other atoms and molecules. This has
been seen only indirectly through spin relaxation
methods\cite{spin0,spin1,spin2,IshikawaYabuzaki} and from the resulting
diffusion\cite{Parniak}.
Our main objective is to observe the random walk behavior of the atoms
themselves directly. 

Perhaps a most direct way to see objects is just to shine light on the
object and observe its absorption or scattering. This is precisely
what is performed in this experiment.  We measure light absorption of
atoms transiting a beam of light.  There are, however, number of
theoretical and technical obstacles that need to be overcome. First,
any observation affects the observed, and it is difficult to directly
observe each collision and the free motion of an atom between
collisions without qualitatively changing their motion, since the
particles performing the measurement have momenta $\sim h/\lambda$,
where $\lambda$ is the mean free path or smaller.  While this is a
fundamental quantum mechanical principle, we can still use photons
with smaller momenta, which have longer wavelengths, to directly
observe atoms undergoing random walk behavior, through their
absorption spectra. The fluctuations in these spectra clearly reflect
the motions of the atoms. Such a direct measurement has not been
performed previously, to our knowledge, and this is what we accomplish
in this work.

Experimentally, light was shone on rubidium atoms in nitrogen buffer
gas.  The light frequency was tuned to a resonance of the rubidium
atoms and the transmitted power of light was measured and its
fluctuations analyzed (\figno{setup}(a)).  While this is, in principle simple, the
fluctuations need to be measured down to orders of magnitude below the
shot noise level, or the standard quantum limit, to uncover the
spectra. 
Shot noise is the quantum statistical noise in the number of detected
photons of the light beam transmitted through the cell, which
contributes to the photocurrent power spectrum as $2eI$ ($e$:
electron charge,  $I$:  photocurrent).
To obtain the spectrum to the desired precision, the experiment is
configured so that statistical analysis involving correlation analysis
is applicable.  To understand the observed behavior, we compute the
spectrum of atoms performing random walks in a light beam
theoretically and derive its form analytically. The theoretical
spectra is compared to the experimentally observed spectra and their
properties are found to be in excellent agreement.

We describe the concept and the setup of the experiment briefly in
\sect{setup}, explain the theory behind the power spectrum of atoms
performing random walks in a gaussian light beam in \sect{theory} and
analyze the experimentally measured spectra in view of the theory in
\sect{exp}. We end with conclusions and discussions
in \sect{disc}.
\section{Design and setup of the experiment}
\label{sec:setup}
\begin{figure}[htbp]\centering
    \includegraphics[width=8.6cm,clip=true]{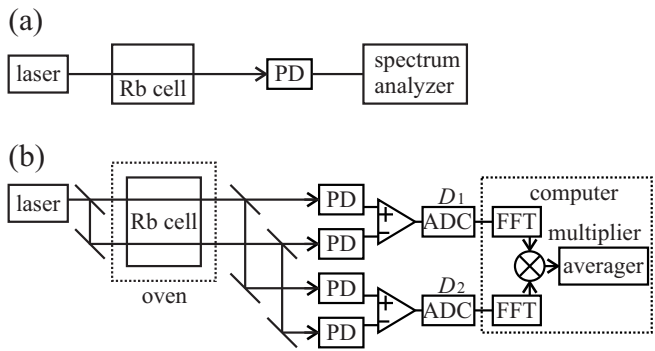}    
    \caption{Configuration of the experiment: resonant laser light
      beam is shone through a rubidium cell, its transmitted power
      measured by the photodetectors (PD) and the power fluctuation
      spectrum is computed.  (a) The conceptual design of the
      experiment. (b) The setup used in the measurements. Differential
      measurement from the two beams is used to remove the correlated
      noise ( $+-$) and the correlations of the transmitted power from
      the same beam, $\cD_{1},\cD_2$  are used to extract the signal (see text). The
      photodetector outputs are converted to a digital signal by the
      analog to digital converters (ADC), Fourier transforms are
      performed (FFT) and then correlations are computed and
      averaged. FFT and averagings are performed on a computer.  }
\label{fig:setup}
\end{figure}
The basic concept underlying our experiment, shown in
\figno{setup}(a), is to just measure the fluctuations in the light
transmitted through atoms and molecules in a cell. In our experiment,
the cells contain rubidium atoms, and nitrogen molecules which serve
as the buffer gas. The obtained fluctuation spectrum is
\begin{equation}
    \label{eq:powerSpectrum}
    S(f) = {1\over\tM}\left|\tilde \cP(f)\right|^2\quad,
    \qquad\tilde \cP(f)=\int_0^\tM\, dt 
    e^{-i2\pi f t}\cP(t)\quad,
\end{equation}
where $\cP$ is the power of the transmitted light and the tilde
denotes its Fourier transform\cite{SpectrumRef}. $f$ is the frequency and $\tM$ is the
measurement time.  In practice, however, this setup in \figno{setup}(a)
by itself is not sufficient to extract the spectrum.  This spectrum
will be almost completely buried under other unwanted noise, in
particular the shot noise.
Often referred to as
the ``standard quantum limit'', shot noise is usually a limiting factor in
the precision of photometric measurements\cite{quantumOptics}.
Therefore, to achieve the desired results, sophisticated noise
reduction methods need to be performed to remove the shot noise, along
with laser noise and the other extraneous noise, to levels which
allows us to recover the fluctuation spectrum.  This full experimental
design is shown in \figno{setup}(b).  Resonant light is shone through
a cell (depth $d_z=44\,$mm, diameter 44\,mm) containing rubidium atoms
in nitrogen buffer gas. The rubidium atoms are at saturation density
and the nitrogen gas has a pressure of 200~Torr.
To reduce the unwanted noise, we compute the correlations of the
measurements from  two independent photodetector measurements
$\cD_1,\cD_2$ of the same atomic vapor. The shot noises in $\cD_{1,2}$
are independent, so that
  \begin{equation}
      \label{eq:corr}
     \vev{\overline{\tilde \cD_1} \tilde \cD_2 }\longrightarrow |\tilde S|^2
     \quad ({\cal N}\rightarrow\infty)
 \quad.
 \end{equation}
where $\vev{\cdots}$ denotes averaging, and the relative statistical error 
here is $1/\sqrt{\cal N}$, with $\cal N$ being  
the number of averagings\cite{am1}. This statistical reduction can
reduce any {\em uncorrelated} noise, including shot noise, to any
desired level, in principle. There is another technical complication
in that unwanted {\em correlated } noise, such as laser noise, also
appears. This was removed through differential measurement, using the
measurements from two light beams, as seen in \figno{setup}(b). The
light beams were separated by 15\,mm in this experiment.  The
averagings of correlations of Fourier transformed transmitted power
measurements, combined with the differential measurements allowed us
to extract the desired spectra, \eqnn{powerSpectrum}.  The incoming
light was tuned to the ${}^{85}$Rb-D$_2$ transition from the hyperfine
levels $5\rm {}^2S_{1/2}$ to $5\rm {}^3P_{3/2}$\cite{RbSpecs} and was
circularly polarized using a quarter-wave plate. The beam radius $w$
was measured with a linear image sensor (Hamamatsu Photonics S9227)
and transmitted power was measured using photodiodes (PD, Hamamatsu
Photonics S5973), as in \figno{setup}.  Light beams with powers, $\cP$,
of $0.1\sim1\,$mW, radius $w=0.3\sim1$\,mm were used and the typical
measurement times were around 2,000\,s.
A similar experimental setup
was used to measure the transit noise, Rabi noise and Zeeman noise of
rubidium atoms without the buffer gas\cite{amRb}, and its principle of
noise reduction is explained in more detail in \cite{amRbExp}.

\begin{figure}[htbp]
    \centering
        \includegraphics[width=8.6cm,clip=true]{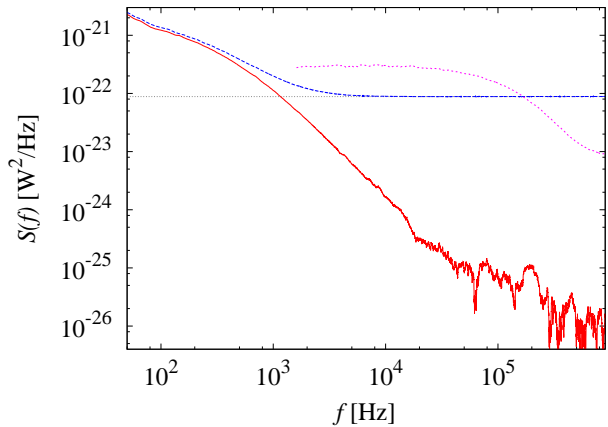}
        \caption{An observed spectrum with and without correlation
          measurements (solid red and dashed blue lines,
          respectively). Without the correlation measurement, the
          spectrum reduces to the shot noise level at higher
          frequencies.  Shot noise level is also shown (grey
          dots). Some extraneous noise still remains at higher
          frequencies. $\cP=58\,\mu$W, $w=0.34\,$mm and temperature
          was $46.4\degC$ in this measurement.  For comparison, the
          transit noise spectrum of the rubidium gas without buffer
          gas with the same $w$ and similar $\cP$ is also shown for
          $f>1.6\,$kHz ($\cP=68\,\mu$W, magenta, short dashes). }
    \label{fig:sn}
\end{figure}

In our experiment, $S$ is generated by atoms transversing the beam,
and shall be referred to as the transit noise below.  The transit
noise spectra are calibrated using the shot noise spectra measured by
the photodetector.
An example of  obtained
spectrum is shown in \figno{sn}, in which the transit noise was
acquired down to four orders of magnitude below the shot noise level.
This spectrum qualitatively differs from that of atoms freely
streaming across the beam, which is also shown.
\section{Spectrum of atoms performing random walks}
\label{sec:theory}
Let us now compute the spectrum of atoms performing random walks in a
gaussian beam of light.
The form of the fluctuation spectrum is determined by the movement of
the individual atoms in a light beam with  inhomogeneous intensity. 
  The electric field strength $\cE(x,y)$ and
the intensity of a monochromatic gaussian light beam, $I(x,y)$ with
the angular frequency $\omega=2\pi f$ are
\begin{equation}
    \label{eq:gaussianBeam}
    \cE(x,y)=\cE_oe^{ikz-i\omega t}\quad,\quad
    I(x,y)=I_oe^{-2(x^2+y^2)/w^2}\quad, 
    \quad I_0={1\over2}c\epsilon_0\cE_0^2\quad,
\end{equation}
where the light beam direction was taken to be along the $z$-axis, $w$
is the beam radius, and $ck=\omega$. $\epsilon_0,c $ are the permittivity
and the speed of light in vacuum. When resonant light is shone on it, 
an atom radiates as a dipole with the power, 
\begin{equation}
    \label{eq:dipole}
    {\wp}(t)={\dip^2\omega\over4\hbar}
    |\cE(x,y)|^2{\Gamma\over(\Delta\omega+kv_z)^2+\Gamma^2/4}\quad.
\end{equation}
Here, $x,y$ is the location of the atom in the $x,y$ plane, $v_z$ is
the velocity of the atom in the beam direction, $\dip$ is the dipole
moment of the atom, $\Delta\omega$ is the amount of detuning and
$\Gamma$ is the line width. In our experiments, the light is tuned to
the resonance and due to the effect of the buffer gas, the line width
is larger than the Doppler shift effect, $kv_z$, so that the dipole
radiation power can be well described by
\begin{equation}
    \label{eq:dip}
    \wp(t)\simeq\sigma I(x,y)\quad,\quad
    \sigma={2\dip^2\omega\over \hbar c\epsilon_0\Gamma}\quad.
\end{equation}
The expression for $\sigma$ is the standard formula for the photon
absorption cross section \cite{Foot}, except for the natural width
being replaced by $\Gamma$.  In this picture, we are treating the
electromagnetic fields semiclassically, which ignores saturation
effects.  Under the conditions of our experiments, the lifetime of
atoms in the excited states is short due to collisions with buffer gas
molecules, making this treatment appropriate.

Since atoms radiate the absorbed light in all directions
independently, they reduce the forward light transmission.  The power
spectrum, \eqnn{powerSpectrum}, for the fluctuations in the
transmission of a gaussian light beam that has passed through atomic
vapor can be obtained by summing over the power radiated by the atoms,
\eqnn{dip}, as
\begin{eqnarray}
    \label{eq:spec}
    S(f)&=&{1\over\tM}\left|\sum_{j}\int_0^\tM\, dt 
      e^{-i\omega t}\sigma I(x_j(t),y_j(t))\right|^2
    \\&=&
    {1\over\tM}\left(\sigma I_0\right)^2 \sum_{j}\int_0^\tM dt\int_0^\tM dt'
    \,e^{-i\omega (t-t')}
    {e^{-2(x_j(t)^2+y_j(t)^2)/w^2}}
    {e^{-2(x_j(t')^2+y_j(t')^2)/w^2}}
    \quad.\nonumber
\end{eqnarray}
Here, the sum is over the atoms, which are  labeled by $j$,
The atoms are performing the random walks in the buffer gas
independently, so that their motions are uncorrelated. This property
was used here.
From here on, we drop the cumbersome index $j$.  
An atom performing a random walk travels as
\begin{equation}
    \label{eq:randomWalk}
    x(t)=x_0+\Delta x(t),\quad 
    \Delta x(t)\equiv \int_{t_0}^tdt'\,\xi(t')\quad,\quad
    x_0=x(t_0)\quad.
\end{equation}
Here $\xi(t)$ satisfies
\begin{equation}
    \label{eq:Wiener}
    \vev{\xi(t)\xi(t')}=2D\delta(t-t')\quad,
\end{equation}
where $D$ is the diffusion constant and $\vev{\cdots}$ denotes
statistical averaging.  Similar relations exist also for $y$.  The sum
of the contributions of atoms can be computed by first averaging over
the random walks, $\Delta x=\Delta x(t),\Delta x' = \Delta x(t')$,
using their probability distribution, $P(\Delta x,\Delta x')$. Since
random walks in each dimension are independent, we may treat spatial
dimensions $x,y$ separately in the averaging. For the $x$-direction,
\begin{equation}
    \label{eq:rWalkX}
    \vev{e^{-{2(x^2(t)+x^2(t'))}/w^2}}=\int d\Delta x\,d\Delta
    x'\,P(\Delta x,\Delta x')\,
    e^{-2\left[(x_0+\Delta x)^2+(x_0+\Delta x')^2\right]/w^2}\quad.
\end{equation}
The distribution is gaussian in
these two variables  with the probability distribution,
\begin{equation}
    \label{eq:prob}
    P(\Delta x,\Delta x')={1\over2\pi|R|^{1/2}}\exp\left[
      -{1\over2|R|}\left(R_{22}\Delta x^2+R_{11}\Delta x'^2
          -2R_{12}\Delta x\Delta x'\right)\right]\quad,
\end{equation}
where  $|R|=R_{11}R_{22}-R_{12}^2$. 
Using the properties of random walks, \eqnn{Wiener}, we derive
\begin{equation}
    \vev{(\Delta x)^2}=2D|t-t_0|=R_{11},\quad
    \vev{(\Delta x')^2}=2D|t'-t_0|=R_{22},\quad
    \vev{\Delta x \Delta x'}=2D\min(|t-t_0|,|t'-t_0|)=R_{12}.
\end{equation}
The integration, \eqnn{rWalkX}, is gaussian, and a straightforward but
cumbersome calculation yields
\begin{equation}
    \label{eq:rWalkX2}
    \vev{e^{-{2(x^2(t)+x^2(t'))}/w^2}}=
    \left(1+4{(R_{11}+R_{22})\over w^2}+16{|R|\over w^4}\right)^{-1/2}
    \exp\left(-4{x_0^2\over w^4}{w^2+2(R_{11}+R_{22})-4R_{12}
        \over 1+4{(R_{11}+R_{22})/ w^2}+16{|R|/ w^4}}\right).
\end{equation}
By combining the $x,y$ directions, replacing the sum over the atoms in
the spectrum, \eqnn{spec} by $n d_z\int dx_0\int dy_0$, where $n$ is
the number density of atoms, we arrive at the following final compact
form for the spectrum.
\begin{equation}
    \label{eq:specFinal}
    S(f)
    ={\pi \over4}nd_z\sigma^2 I_0^2w^2
    \int_{-\infty}^\infty    d\tau\,{e^{-i\omega\tau}\over1+4D|\tau|/w^2}
    ={\pi nd_z\sigma^2 I_0^2 w^4\over 8D}\Re \left[-e^{i\omega'}{\rm
      Ei}(-i\omega')\right]\quad.
\end{equation}
Here we let $t\longrightarrow-\infty$, ${\rm Ei}(x)$ is the
exponential integral function\cite{Gradshteyn} and $ \omega'=\omega
{w^2/(4D)}$. 
$1/\tM$ factor cancels one of the integrals in the formula for the
spectrum, \eqnn{spec}, since the integrand depends only on the time
difference $t'-t$.
This formula, \eqnn{specFinal}, is the transmission power spectrum of
atoms individually performing random walks in a gaussian light
beam. Since
\begin{equation}
    \label{eq:EiExpansion}
    \Re \left[-e^{i\omega'}{\rm
        Ei}(-i\omega')\right]={1\over\omega'^2}
    -{6\over\omega'^4}+{\cal O}\left({1\over\omega'^{6}}\right)\quad,    
\end{equation}
we obtain the high frequency behavior of the spectrum as
\begin{equation}
    \label{eq:specFExp}
    S(f) ={C(w,n,\cP)\over f^2}+\cdots    ,\quad
    C(w,n,\cP)={2n\sigma^2d_z{\cal P}^2D\over\pi^3 w^4}\quad.
\end{equation}
The theoretical spectrum, \eqnn{specFinal}, is uniquely determined
from the properties of atoms and the experimental parameters.  It
should be noted that the diffusion constant $D$ appears in the
spectrum, directly reflecting the motion of atoms. The shape of the
spectrum is governed by $D$ and $w$, while the overall coefficient
further depends on $\cP,n$
 While the spectrum was derived with atoms in mind, it should be
 evident that the spectrum is applicable to any light absorbing
 particles performing random walks in a gaussian beam.
\section{Observed spectra of atoms in buffer gas}
\label{sec:exp}
We shall now compare the properties of the spectrum derived in the
previous section with the spectra observed experimentally, as
explained in \sect{setup}.  We first analyze the properties of the
system in our experimental situation, partly to understand concretely
the background for the approximations made in the previous section.
Rubidium atoms are in nitrogen buffer gas at 200\,Torr at temperatures
of 40 to 50$\degC$. The average velocity of a rubidium atom is
300\,m/s, the Doppler width is $320$\,MHz and the width of the
rubidium atoms in nitrogen buffer gas is $\Gamma/(2\pi)=3.7\,$GHz
\cite{RbBuffer}, which is about 600 times larger than the natural
width\cite{RbSpecs}.
In the buffer gas, the width of the rubidium resonance
is
larger than the hyperfine level splittings, thereby reducing the
system to a two level system to a good approximation. Due to
collisional broadening, $kv_z/(2\Gamma) 
\sim0.04$ so that Doppler effects are small in
\eqnn{dipole}, as mentioned above.  The ratio of photon momentum to
the average rubidium atom momentum is about $3\times10^{-5}$ making the
measurement essentially passive. 
Light beams with powers $\cP\lesssim2$\,mW and beam radii ($w$) of
0.2 to 1\,mm were used.
The diffusion constant for the rubidium atoms in nitrogen buffer gas
at 200\,Torr is $D=1.59(4)\times10^{-5}\,{\rm m^2/s}$
\cite{IshikawaYabuzaki}.
Then, the transit time for an atom across a light beam with $w=1\,$mm
is around 4\,ms. The average number of photons absorbed by an atom
during this transit is $\cP \sigma/(2h\nu D)$, which is
interestingly independent of the beam size. 
This number is around $2\times10^3$, for $\cP = 1\,$mW under
our experimental conditions.  Due to the short excited state life
time, the ratio of atoms in the excited state is $(\dip \cE/\hbar
\Gamma)^2 \sim2\times10^{-5}$ for $\cP=1$\,mW, $w=1\,$mm, so that 
saturation effects should be negligible in our measurements.

\begin{figure}[htbp]
    \centering
    \includegraphics[width=8.6cm,clip=true]{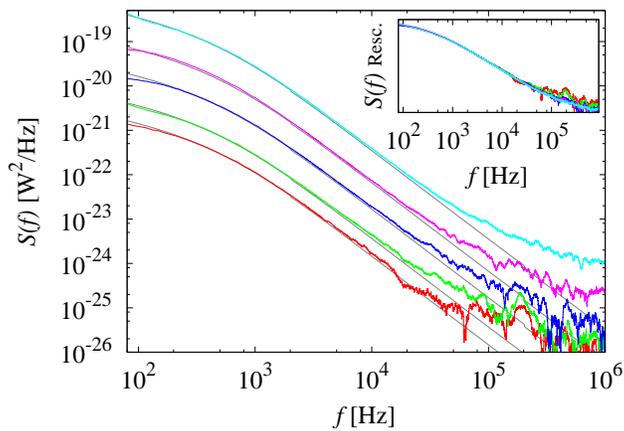}
    \caption{Power spectra of rubidium atoms in nitrogen gas
      (200\,Torr) and the corresponding theoretical spectra (thin grey),
      \eqnn{specFinal}. The spectra have larger magnitudes for larger
      light power $\cP$. The theoretical and experimental spectra agree
      well. (Inset) The experimental spectra rescaled  by their overall
      coefficient, which show that the shapes of the spectra are the
      same to a high degree. The vertical scale is denoted as 
      ``$S(f)$ Resc.'' and the units on the vertical scale is arbitrary.
      $\cP= 58.0$ (red), 92.8 (green), 197 (blue), 429 (magenta), 912
      (cyan)\, $\mu$W. $w=0.96$\,mm and the temperature of the gas is
      $46.4\degC$. Larger $\cP$ leads to a larger signal. }
    \label{fig:bufComp}
\end{figure}

There are several distinct properties of the theoretical spectrum,
\eqnn{specFinal};
\begin{enumerate}
  \item the shape of the spectrum is independent of $\cal P$.
  \item $S(f)\sim1/f^2$ for $f\gg D/w^2$, but has different  behavior at
    lower frequencies, so that  $D$ can be extracted from the spectrum.
  \item $S(f)$ is proportional to $\cP^2$ and $w^{-4}$.
  \item $S(f)$ is proportional to $n\sigma^2$, from which  $n\sigma^2$
    can be measured.
\end{enumerate}
We shall now investigate these properties in the experimental
results: While some physical properties of atoms were provided above
as a background, we shall use only the quantities measured in our
experiments, $\cP, w, d_z$ and the experimental spectra in this
analysis, unless noted otherwise.  The transit noise spectra for the
buffered rubidium gas is shown in \figno{bufComp} for powers $\cP$
varying more than over an order of magnitude. One overall coefficient,
$C(w,n,\cP)$, has been extracted from each experimental spectrum by
fitting to $1/f^2$ at higher frequencies.  From the shape of the
measured spectra, $D=6.0\times10^{-5}\,\rm m^2/s$ was obtained.  The
theoretical spectra are compared against the experimentally measured
ones in \figno{bufComp} and they are all seen to agree quite well. $D$
obtained here is quite consistent with the previously measured
values
\cite{IshikawaYabuzaki}. The measured spectra divided by the
overall coefficient $C(w,n,\cP)$ are shown in
\figno{bufComp}(inset). Theoretically, the rescaled spectra should be
identical, which they are, to a high degree.

\begin{figure}[htbp]
    \centering
    \includegraphics[width=8.6cm,clip=true]{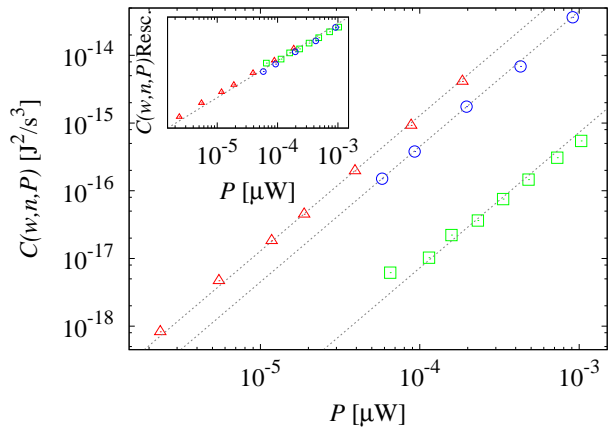}
    \caption{Power, $\cP$, dependence of $C(w,n,\cP)$: $C(w,n,\cP)$
      and fits to them proportional to $\cP^2$ (grey dashes) are
      plotted for measurements at temperatures and $w$ values
      $(44.6\degC, 0.96\,{\rm mm})\ (\Box)$ , $(46.4\degC, 0.34\,{\rm
        mm}) (\circ)$, and $(50.0\degC, 0.34\,{\rm mm})
      (\triangle)$. (Inset): $\cP$ dependence of
      $C(w,n,\cP)w^4/n_{\rm theory}$ for the same spectra and its fit
      to $\cP^2$  (grey dashes). The data can be seen to fall on a
      single line. (Vertical scale  labeled as `` $C(w,n,\cP)$ Resc."
      and arbitrary units used.)    }
    \label{fig:bufPComp}
\end{figure}
The dependence of $C(w,n,\cP)$ on $\cP$ is shown in
\figno{bufPComp}. The $\cP^2$ behavior is shown and clearly fits the
experimental results quite well. This behavior, which shows no
saturation effects, is consistent with the short life time of the
excited state, as mentioned above.  The spectrum, \eqnn{specFinal}, is
proportional to $n/w^{4}$ and so that if they are rescaled its
inverse, they should all agree. It can be seen that this is indeed the
case in \figno{bufPComp}~(inset). For this rescaling, the number
densities, $n_{\rm theory}$ from the literature\cite{RbSpecs} were used.
These properties agree with those of the theoretical spectrum,
\eqnn{specFinal}, and show that the coefficient $C(w,n,\cP)$, is a
function of $\cE_0$, for a gaussian beam.  It should be recalled,
however, the shape of the spectrum is rather governed by $w$ and $D$.

\begin{figure}[htbp]
    \centering
    \includegraphics[width=8.6cm,clip=true]{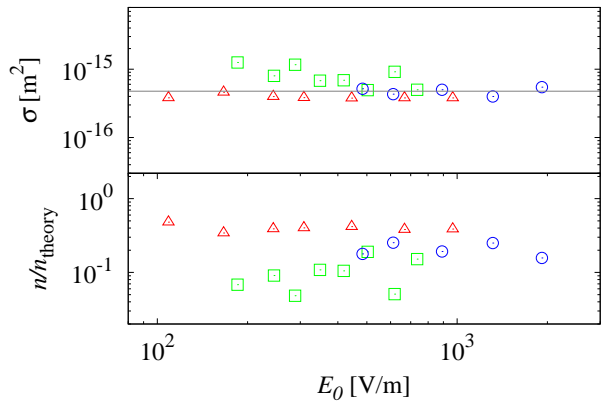}
    \caption{Cross section, $\sigma$, and the rubidium number density,
      $n$, extracted from the spectra used in \figno{bufPComp} (same
      symbols used). (Upper panel) Measured $\sigma$. The theoretical
      cross section, \eqnn{dip}, is shown (grey dashes). (Lower panel)
      Measured rubidium number density, $n$, relative to the number
      density in the literature ($n_{\rm theory}$). }
    \label{fig:bufN}
\end{figure}

Since the spectrum, \eqnn{specFinal} is uniquely determined from the
properties of the atoms and the experimental conditions, both $n$ and
$\sigma$ can be extracted from the results as follows.  By applying
Beer's Law\cite{Foot} for the transmission rate, $\exp(-n\sigma d_z)$,
we can extract $n\sigma$ for each experiment using the measured
transmission rate.  Then, $n\sigma^2$ can be obtained from
$C(w,n,\cP)$ in from experimentally measured physical parameters.
Combining these data, we arrive and $\sigma, n$ values extracted from
each set of spectrum and transmission rate measurement, which we show
in \figno{bufN}. The measured cross section is compared to the
theoretical formula in \eqnn{dip}, using the line widths in
\cite{RbBuffer} and the density, $n$, is compared to the theoretical
rubidium vapor saturation density\cite{RbSpecs}. $\cP$, $w$ and the
temperature were measured independently of the spectra and the
theoretical and experimental results agree reasonably.  Compared to
the relative properties analyzed above, the absolute values of the
spectra are much more sensitive to various uncertainties, both
experimental and theoretical.

$n$, $\sigma$ are seen to be both independent of $\cE_0$ within
experimental uncertainties, as would be expected for non-saturated
vapor. The photon absorption cross section of atoms in buffer gas has
not been directly measured, to our knowledge, and the simple formula
\eqnn{dip} seems to be a good approximation.
While the system should be effectively a two level system due to collisional
broadening, more detailed analysis may be necessary to establish the
numerical factors precisely.
Some of the more significant experimental uncertainties also should be
mentioned: The power of the light beam entering the cell and
transmitted light can differ by up to 60\,\%, and the intensity of the
beam changes along the beam itself. In this work, we have consistently
used the transmitted power of light measured by the photodetectors
(\figno{setup}(b)) when referring to the power of the light beam,
$\cP$. Furthermore, the incoming and the transmitted light beam are
not perfectly gaussian and the temperature of the gas inside the cell,
which affects the number density of atoms, can have uncertainties of a
few degrees throughout the whole measurement.
\section{Discussion}
\label{sec:disc}
In this work, the time dependent fluctuations of the light transmission
power of atoms in buffer gas were experimentally measured, down to
four orders of magnitudes below the shot noise level in their
spectra. On the theoretical side, the spectrum of atoms performing
random walks in buffer gas was investigated and an analytic form of
the spectrum was derived. The theoretical spectrum was found to
describe the observed spectra quite well.

From a measured spectrum and the transmission rate, the diffusion
constant, $D$, the number density of the rubidium atoms, $n$, and the
photon absorption cross section of the atom in buffer gas, $\sigma$,
and can all be obtained. Namely, we can obtain $D$ from the shape of
the spectrum, $n\sigma$ from the absorption rate, and then $n\sigma^2$
from the overall size of the spectrum. In typical experimental
situations, the combination $n\sigma$ can be measured from absorption,
but not $n,\sigma$ separately. This is a distinct feature of the
fluctuation measurement and we believe that it is interesting and
practical to be able to measure these quantities at the same time, in
which the same experimental conditions are guaranteed.

In the theory of the spectrum of atoms performing random walks, we
used a semiclassical picture. This seems to us to be the simplest
approach and one that should be applied first. It also seems to
describe the observed spectra remarkably well.  Since the Doppler
width is much smaller than the line width and the monochromatic light
is tuned to the resonance, the assumption that atoms are in resonance
is reasonable. It was tacitly assumed that $\sigma^2$ in the
spectrum, \eqnn{specFinal}, is the square of $\sigma$ found in the
average transmission rate.  However, the time scale for de-excitation
is $3\times10^{-10}$\,s and the decay time is probabilistic for
individual atoms. The cross sections used in the absorption rate and
in the spectrum, \eqnn{specFinal}, should therefore be regarded as
averaged values.  Furthermore, this time scale is a factor of $15$
smaller from the time scale obtained from the diffusion constant in
\cite{IshikawaYabuzaki}, $6D/v^2\sim4\times10^{-9}$\,s ($v^2$ is the
average velocity squared of rubidium atoms in three spatial
dimensions). While the de-excitation time scale needs not be identical
to the diffusion time scale, they both come from atoms colliding with
buffer gas molecules.  It is possible that there are corrections,
perhaps including quantum electrodynamics contributions, to the
semiclassical picture presented above, which would be of considerable
interest. We believe that the transit noise measurements could
provide a way to address these questions in more detail.
And, we find it fascinating that we can make direct observations of
atoms performing random walks.

Since the method we use to measure the transit noise of atoms  has
not been used previously, it allows us to investigate the validity of
the standard basic physics principles, which we feel is important and
was also done here.  Furthermore, a new method of measurement can open
up new approaches to understanding atom photon interactions. We expect
this conceptually simple method to bring about further insight into
their properties. Our approach of measuring transit noise is
applicable to any atoms or molecules, regardless of the density, pure
or mixed, as long as a resonant light source is used.  In particular,
properties of atoms in buffer gas and rubidium atoms have played an
important role in various active areas of fundamental fields of study
in physics, such as atomic clocks\cite{atomicClock,atomicClockRev},
Bose Einstein condensation of cold atoms\cite{BEC}, as well as some
applications to other fields\cite{RbMed}.  Methods to analyze the
properties of atom photon interactions is of interest also from such
considerations.

\acknowledgments 
K.A. was supported in part by the Grant--in--Aid for
Scientific Research (\#15K05217) from the Ministry of Education,
Culture, Sports, Science and Technology of Japan, and a grant from Keio
University.

\end{document}